\documentclass[12pt,preprint]{aastex}

\usepackage{apjfonts}
\usepackage{amsmath}
\usepackage{epsfig}
\usepackage{amssymb}

\renewenvironment{figure}{\begin{figure*} }{\end{figure*}}

\newcommand{\bu}{{\bf u}}

\newcommand{\bj}{{\bf j}}
\newcommand{\grad}{{\mathbf \nabla}}
\renewcommand{\div}{{\mathbf \nabla} \cdot}
\newcommand{\curl}{{\mathbf \nabla} \times}

\newcommand{\ephi}{{ \hat{\bf e}_{\phi}}}

\newcommand{\dd}{{\rm d}}
\newcommand{\ro}{r_{\rm o}}
\newcommand{\deltao}{\delta_{\rm o}}
\newcommand{\thetao}{\theta_{\rm o}}
\newcommand{\ri}{r_{\rm i}}
\newcommand{\deltai}{\delta_{\rm i}}
\newcommand{\thetai}{\theta_{\rm i}}
\newcommand{\bBp}{{\bf B}_{\rm p}}
\newcommand{\bB}{{\bf B}}

\begin{document}

\title{The rotation rate of the solar radiative zone}

\author{P. Garaud$^1$ \& C. Guervilly$^2$} 

\affil{$^{1}$ Department of Applied Mathematics and Statistics, Baskin School of Engineering, University of California Santa Cruz, Santa Cruz, USA. \\
$^{2}$ Laboratoire de G\'eophysique Interne et Tectonophysique, CNRS, Universit\'e Joseph-Fourier, Grenoble, France.
}

\maketitle

\vspace{0.5cm}


\centerline {\it Abstract} 

\vspace{0.2cm}

The rotation rate of the solar radiative zone is an important diagnostic
for angular-momentum transport in the tachocline and below. 
In this paper we study the contribution of viscous and magnetic 
stresses to the global 
angular-momentum balance. By considering a simple linearized 
toy model, we discuss the effects of field geometry 
and applied boundary conditions on the predicted rotation profile and 
rotation rate of the radiative interior. We compare these analytical 
predictions with fully nonlinear simulations of the dynamics of the 
radiative interior, as well as with observations. We discuss the implications
of these results as constraints on models of the solar interior. 


\vspace{0.5cm}

\section{Introduction}

Helioseismic inference of the rotation profile of the solar interior
has revealed two spatially-distinct regions: an outer 
differentially-rotating shell surrounding an inner uniformly
rotating core (Christensen-Dalsgaard \& Schou, 1988;
Kosovichev, 1988; Brown {\it et al.} 1989; Dziembowski {\it et al.} 1989). 
The transition between the two regions, the solar 
tachocline, is located precisely at the base of the solar convection zone.
It is surprisingly sharp with an average width no 
larger than a few percent of the solar radius (Charbonneau {\it et al.} 1999, 
Elliott \& Gough, 1999).
Our understanding of this peculiar rotation profile 
has steadily marched on in the past three decades, benefiting greatly 
from the high-performance computing revolution. 
Today, the field is ripe for more quantitative comparisons between
models and observations, and  has begun focusing on specific 
reference points, such 
as the overall pole-to-equator difference in the rotation rate, the inclination
of the isorotation contours, the thickness of the tachocline and finally, the 
subject of this paper, the rotation rate of the radiative interior.

If one assumes that the solar interior is in a dynamically quasi-steady
state, then the rotation rate of the radiative zone $\Omega_{\rm rz}$ 
can be thought of as a weighted average of the rotation profile observed 
near the base of the convection zone:
\begin{equation}
\Omega_{\rm cz}(\theta) \simeq \Omega_{\rm eq} (1-a_2 \cos^2\theta
- a_4 \cos^4\theta) \mbox{   ,   }
\label{eq:ocz}
\end{equation}
where $\Omega_{\rm eq}/2\pi
= 463 {\rm nHz}$, $a_2 = 0.17$ and $a_4  = 0.08$ (Schou {\it et al.}
1998; Gough 2007). Hence we can formally write 
\begin{equation}
\Omega_{\rm rz} = \int_{0}^{\pi/2} W(\theta) \Omega_{\rm cz}(\theta) \sin\theta \dd \theta  \mbox{   ,   }
\end{equation}
where the weight function $W(\theta)$ depends uniquely on the nature of 
angular-momentum transport in the tachocline. Observations provide us 
with relatively precise measurements of $\Omega_{\rm rz}$ 
(Schou {\it et al.} 1998), with 
\begin{equation}
\Omega_{\rm rz}/2\pi = 430 {\rm nHz} \rightarrow \Omega_{\rm rz} \simeq 0.93 \Omega_{\rm eq}  \mbox{   .   }
\label{eq:orz}
\end{equation}
Can this information be used to constrain theoretical models of the 
solar interior?

If angular-momentum transport in the radiative zone and the tachocline 
were purely viscous, 
the rotation rate of the deeper regions 
($r \rightarrow$ 0) would be the same as that of the mean specific angular 
momentum of the convection zone (Gilman, Morrow \& DeLuca, 1989), 
or in other words 
$W_{\rm visc}(\theta) = \sin^2\theta$ so that
\begin{equation}
\Omega_{\rm visc} = \left(1 - \frac{a_2}{5} - \frac{3a_4}{35} \right) \Omega_{\rm eq} \simeq 0.959 \Omega_{\rm eq} \mbox{   .   }
\label{eq:ovisc}
\end{equation}
Of course, purely viscous transport cannot account for the observed 
uniform rotation. Spiegel \& Zahn (1992) proposed the first 
tachocline model, in which angular momentum is primarily transported 
by anisotropic turbulence. Since the stratification of the 
radiative zone strongly inhibits radial fluid motions, they modelled 
the effects of this turbulence as a dominantly horizontal diffusion process, 
and found that the interior would indeed relax to uniform 
rotation beneath the tachocline with an angular velocity
\begin{equation}
\Omega_{\rm SZ} = \left(1 - \frac{3a_2}{7} - \frac{5a_4}{21} \right) \Omega_{\rm eq} \simeq 0.908 \Omega_{\rm eq} \mbox{   .   }
\end{equation}
Gough \& McIntyre (1998, GM98 hereafter) later argued 
against this model on the ground that
two-dimensional turbulence does not act to diffuse angular momentum 
horizontally (e.g. Tobias, Diamond \& Hughes 2007). 
Moreover, the predicted value
$\Omega_{\rm SZ}$ is sufficiently far from the observed value to be 
ruled out by the observations. 

In the past decade, magnetized models have become
more widely accepted as the simplest explanation for the observed uniform 
rotation of the radiative zone (R\"udiger \& Kitchatinov 1997; GM98; 
see Garaud, 2007 for a review).
A large-scale primordial magnetic field, strictly {\it confined} beneath 
the convection zone can indeed robustly maintain a state of uniform 
rotation through Ferraro's law of isorotation (Ferraro, 1937). The  
confinement of the primordial field is thought to result from interactions with
large-scale meridional flows originating from the convection zone, as 
originally proposed by GM98, and first shown 
numerically by Garaud \& Garaud (2008) (GG08 hereafter). 

However, little attention 
has been given so far to the predicted rotation rate of the radiative zone 
in these models. The linear simulations of R\"udiger \& Kitchatinov (1997),
which assume a given confined poloidal field structure, suggest that 
$\Omega_{\rm rz} \simeq 0.97 \Omega_{\rm eq}$, a value which is much 
larger than observations. The nonlinear simulations of GG08 
on the other hand suggest that $\Omega_{\rm rz} \simeq 0.87 
\Omega_{\rm eq}$, a value which is far too low. Can we understand these
numerically determined values in terms of simple force-balance arguments? 
We show in this paper that it is indeed possible. Moreover, 
much can be learned from this exercise in terms of relating models 
to the real Sun. 

We begin in \S2 by describing a linearized toy model 
of the radiative zone in which the poloidal component of the field is 
fixed (R\"udiger \& Kitchatinov 1997; MacGregor \& 
Charbonneau 1999). As in the work of MacGregor \& Charbonneau (1999),
we study two different field geometries: an open dipole (\S\ref{sec:open}) 
and a confined dipole (\S\ref{sec:closed}). In both cases 
we study the effect of boundary conditions on the predicted rotation 
profile and rotation rate of the radiative interior. 
We discuss the implications of our findings by comparing these toy-model 
predictions with fully nonlinear simulations of the dynamics of the 
radiative interior in \S\ref{sec:disc}, and conclude in \S6.

\section{A toy model of angular momentum transport in the solar radiative zone}
\label{sec:model}

Let us consider the simplest possible setup in which to study the interaction 
between large-scale fields and flows in a spherical shell: the homogeneous 
magnetized spherical Couette flow. We thus consider a spherical shell 
containing a homogeneous incompressible conducting fluid; 
the outer radius of the shell is $\ro$ while the inner radius 
is $\ri$. The uniform density of the fluid is $\rho$, its viscosity $\nu$ 
and magnetic diffusivity $\eta$. We model the medium outside 
the spherical shell as a solid, and allow this solid to 
have various conducting properties.
The inner core, for radii $r < \ri$, is assumed to be permeated by 
currents which maintain a poloidal dipolar magnetic field. 

This setup has been extensively studied in the 
geophysical literature as a model of the Earth's interior and is also
used as a basis for studying spherical Couette flow dynamo experiments 
(e.g. 
Dormy, Cardin \& Jault 1998; Hollerbach, 2000; Dormy, Jault \& Soward 2002; 
Hollerbach, Canet \& Fournier, 2007). It has also been used in the 
solar context by R\"udiger \& Kitchatinov (1997) and MacGregor \& 
Charbonneau (1999). As in these latter papers, we 
neglect the meridional flows entirely. 
The validity of this approximation is discussed in 
\S\ref{sec:disc}. The flow considered is therefore defined in the spherical 
coordinate system $(r,\theta,\phi)$ as
\begin{equation}
\bu = \left(0,0, r\sin\theta \Omega(r,\theta)\right)\mbox{   .   }
\end{equation}
As a result of this assumption, the fluid within the spherical shell does 
not influence the imposed poloidal field. 

We introduce the flux function $A(r,\theta)$ such that
\begin{equation}
\bBp = \curl \left( \frac{A}{r\sin\theta} \ephi \right)\mbox{   .   }
\end{equation}
The toroidal component of the magnetic field 
$B_\phi$ is generated by the $\Omega$-effect 
induced by the azimuthal flow. For simplicity 
of notation, we introduce the new variable $S(r,\theta)$ such that
\begin{equation}
S = r\sin\theta B_\phi \mbox{   ,   }
\end{equation} 
so that the total magnetic field can 
be written as 
\begin{equation}
\bB = \frac{1}{r\sin\theta} \left( \frac{1}{r} \frac{\partial A}{\partial \theta}, - \frac{\partial A}{\partial r}, S \right) \mbox{   .   }
\end{equation}

The dynamics of this reduced 
system are entirely described by the azimuthal component of the momentum 
equation as well as the azimuthal component of the induction equation. These
are expressed in the spherical coordinate system as (cf. R\"udiger \& Kitchatinov 1997):
\begin{eqnarray}
&& \eta \left[ \frac{\partial}{\partial \theta} \left( \frac{1}{r\sin \theta} \frac{\partial S}{\partial \theta} \right) 
+ \frac{r}{\sin\theta} \frac{\partial^2 S}{\partial r^2} \right] 
= 
r \left( \frac{\partial \Omega}{\partial \theta} \frac{\partial A}{\partial r} 
- \frac{\partial \Omega}{\partial r} \frac{\partial A}{\partial \theta} \right) ,
\label{eq:indphi}
\\
&& \rho \nu \left[ \frac{1}{\sin^3 \theta} \frac{\partial}{\partial \theta} \left( \sin^3 \theta \frac{\partial \Omega}{\partial \theta} \right)
+ \frac{1}{r^2} \frac{\partial}{\partial r} \left( r^4 \frac{\partial \Omega}{\partial r} \right) \right] 
= \ 
\nonumber
\\
& & \qquad \qquad \qquad \frac{1}{4\pi r^2 \sin^3 \theta} \left( \frac{\partial A}{\partial r} \frac{\partial S}{\partial \theta}
- \frac{\partial A}{\partial \theta} \frac{\partial S}{\partial r} \right) .
\label{eq:NSphi}
\end{eqnarray}

Note that if both diffusion terms are neglected, then the equations reduce to
\begin{eqnarray}
\nabla \Omega \times \nabla A \propto \bBp \cdot \nabla \Omega = 0 \mbox{   ,   } \label{eq:ferraro} \\
\nabla S \times \nabla A \propto \bBp \cdot \nabla S = 0 \mbox{   .   } \label{eq:forcefree}
\end{eqnarray}
The respective solutions of these equations are simple: $\Omega$ and $S$ 
must be constant on poloidal magnetic field lines 
(which are lines of constant $A$). Equation (\ref{eq:ferraro}) is an expression 
of Ferraro's isorotation theorem, while (\ref{eq:forcefree}) expresses the fact that in this steady-state axisymmetric system the azimuthal component of 
the Lorentz force must be zero. 

The boundary conditions are selected as follows. We 
consider that the inner boundary is rotating uniformly with
\begin{equation}
\Omega(\ri,\theta) = \Omega_{\rm in} \mbox{   ,   }
\label{eq:ominbc}
\end{equation}
while the outer boundary is rotating differentially as 
\begin{equation}
\Omega(\ro,\theta) = \Omega_{\rm cz} (\theta)\mbox{   .   }
\label{eq:omczbc}
\end{equation}
The angular velocity $\Omega_{\rm in}$ is selected, as 
in Garaud (2002) and GG08, in such a way as to guarantee 
that the total torque applied to the system is zero. Hence, we require 
that 
\begin{equation}
\int_{0}^{\pi/2} \left( \rho \nu r^2 \sin^2 \theta \frac{\partial \Omega}{\partial r} + r \sin \theta \frac{B_r B_{\phi}}{4\pi } \right) \sin \theta \dd \theta =0 \mbox{   .   }
\label{eq:notorque}
\end{equation}
Note that in this steady-state calculation, equation (\ref{eq:notorque}) only
needs to be applied at one particular radius $r$ to be valid everywhere. 
We now calculate $\Omega_{\rm in}$ for a variety of poloidal field configurations and boundary conditions on the azimuthal magnetic field. 

\section{Solution for an open field configuration}
\label{sec:open}

In order to represent an open field configuration, we select 
\begin{equation}
A(r,\theta) = \frac{B_0}{2} \ri^3 \frac{\sin^2\theta}{r}  \mbox{   .   }
\label{eq:Aopen}
\end{equation}
The normalizing constant is chosen so that this flux function represents
the poloidal magnetic field $\bBp$ which is the exact axisymmetric 
solution of the equation $\grad^2 \bBp = 0$ 
in the whole space, matches to a point dipole at $r=0$ and 
decays as $r\rightarrow \infty$, and finally, which has amplitude 
$B_0$ on the polar axis at radius $r= \ri$. 

\begin{figure}[h]
\epsscale{0.3}
\plotone{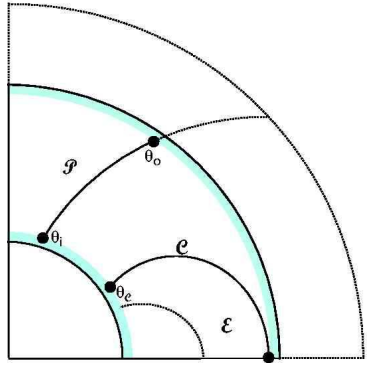}
\caption{Unconfined field geometry, showing the polar region (${\cal P}$) 
and the equatorial region  $({\cal E})$ separated by the last connected 
field line $({\cal C})$.}
\label{fig:opensetup}
\end{figure}

Dormy, Cardin \& Jault (1998) and Dormy, Jault \& Soward (2002) presented
the first analytical studies of the linear dynamics of magnetized 
spherical Couette flows. The following analysis is analogous to 
their approach in the limit where meridional flows are neglected, but 
investigates the case of a differentially rotating outer boundary. 
Following their results we expect the presence of two 
boundary layers near the inner and outer boundaries respectively, as
well as an internal shear layer along the ``last connected field line'', 
as shown in Figure \ref{fig:opensetup}. 
This particular field line $({\cal C})$  separates the equatorial 
region  $({\cal E})$ from the polar region  $({\cal P})$. 

In the equatorial region $({\cal E})$, 
every field line (${\cal F}$) originating from the 
inner core in the Northern hemisphere re-enters the core at a symmetric
latitude in the Southern hemisphere. In the limit of negligible diffusion, 
the angular velocity must be constant along (${\cal F}$) implying that
the entire equatorial region must rotate with angular velocity 
$\Omega_{\rm in}$. The function $S$ must also be constant
along (${\cal F}$), but in addition is antisymmetric with
respect to the equator. The only possibility is therefore $S=0$ everywhere
in $({\cal E})$. 

In the polar region $({\cal P})$, field lines originating from 
a co-latitude $\thetai$ on the inner core extend out to co-latitude 
$\thetao$ on the outer boundary (see Figure 1), where
\begin{equation}
\frac{\sin^2 \thetai}{\ri}= \frac{\sin^2 \thetao}{\ro} \mbox{   .   }
\label{eq:thetaio}
\end{equation}
The field line $\cal{C}$ emerges from the inner core at 
co-latitude $\theta_{\cal C}$, with
\begin{equation}
\sin \theta_{\cal C}= \left(\frac{\ri}{\ro}\right)^{1/2} \mbox{   .   }
\end{equation} 

We now consider the solutions of the problem under various types of boundary
conditions for the magnetic field.

\subsection{Vanishing toroidal field on the boundaries}
\label{sec:insulating}

We first consider the case where $S = r\sin\theta B_\phi$ is required
to be zero both at the inner and the outer boundary:
\begin{equation}
S(\ri,\theta) = S(\ro,\theta) = 0
\label{eq:s0bc}
\end{equation}
These can be thought of as ``insulating'' boundary conditions.

\subsubsection{Analytical solutions}

To study the boundary layer near $\ro$ we introduce the 
scaled variable 
\begin{equation}
  \xi = \frac{\ro-r}{\deltao f(\theta)} \mbox{   ,   }
\label{eq:xidef}
\end{equation}
where the typical boundary layer thickness $\deltao$ and its form function $f(\theta)$ both remain to be determined. We assume 
(and later verify) that for low diffusivities $\deltao$ is very small 
compared with the global scales in the system ($\ri,\ro$). Substituting 
this new variable in the governing equations (\ref{eq:indphi}) and 
(\ref{eq:NSphi}), using the chain rule 
\begin{equation}
\frac{\partial}{\partial r} \rightarrow -\frac{1}{\deltao f(\theta)} \frac{\partial}{ \partial \xi} \mbox{  and  } \frac{\partial}{\partial \theta} \rightarrow \frac{\partial}{\partial \theta} - \xi\frac{f'(\theta)}{f(\theta)} \frac{\partial }{\partial \xi}
\end{equation}
and keeping only the lowest order terms in $\deltao$ yields
\begin{eqnarray}
\frac{\partial^2 S}{\partial \xi^2}
&=&  B_0 \frac{\ri^3 \deltao f(\theta)}{\eta \ro} \sin^2\theta \cos\theta  \frac{\partial \Omega}{\partial \xi} \mbox{   ,   } \nonumber \\ 
 \frac{\partial^2 \Omega}{\partial \xi^2} 
&=& \frac{B_0}{4\pi \rho \nu  } \frac{\ri^3 \deltao f(\theta)}{\ro^5} \frac{\cos\theta}{\sin^2\theta }  \frac{\partial S}{\partial \xi} \mbox{   .  } 
\end{eqnarray}
Combining the two equations yields
\begin{equation}
\frac{\partial^3 S}{\partial \xi^3}
=  \frac{\partial S}{\partial \xi}  \mbox{    and   }  
 \frac{\partial^3 \Omega}{\partial \xi^3} = \frac{\partial \Omega}{\partial \xi}  \mbox{   ,  } 
\label{eq:SOmeqs}
\end{equation}
provided we define
\begin{equation}
\deltao^2 = \frac{4\pi \rho \nu \eta }{B^2_0} \frac{\ro^6}{\ri^6}  \mbox{  and   } f(\theta) = \frac{1}{\cos\theta} \mbox{   .  } 
\end{equation}
The solutions to this set of equations which remain bounded as $\xi\rightarrow + \infty$ are
\begin{eqnarray}
&& S_{\rm o}(\xi,\theta) = s_{\rm o}^{(0)}(\theta) + s_{\rm o}^{(1)}(\theta) e^{-\xi}\mbox{   ,  }  \nonumber \\
&& \Omega_{\rm o}(\xi,\theta) = \omega_{\rm o}^{(0)}(\theta) + \omega_{\rm o}^{(1)}(\theta) e^{-\xi} \mbox{   ,  } 
\end{eqnarray}
where the index ``o'' denotes that this solution is only valid in the outer boundary layer. The integrating functions $s_{\rm o}^{(1)}(\theta)$ and $\omega_{\rm o}^{(1)}(\theta)$ are related to one another by the equation
\begin{equation}
s_{\rm o}^{(1)}(\theta) = - \left(\frac{4\pi \rho\nu}{\eta}\right)^{1/2} \ro^2 \sin^2\theta  \omega_{\rm o}^{(1)}(\theta) \mbox{   .  } 
\label{eq:rel1}
\end{equation}

A very similar calculation can be done near $\ri$ introducing 
the scaled variable 
\begin{equation}
\zeta  = \frac{r - \ri}{\deltai f(\theta)} \mbox{   ,  }
\label{eq:zetadef}
\end{equation}
with
\begin{equation}
\deltai^2 = \frac{4\pi \rho \nu \eta }{B^2_0 }  \mbox{   ,   } 
\end{equation}
yielding similar governing equations (see (\ref{eq:SOmeqs})) and therefore 
the solutions 
\begin{eqnarray}
&& S_{\rm i}(\zeta,\theta) = s_{\rm i}^{(0)}(\theta) + s_{\rm i}^{(1)}(\theta) e^{-\zeta} \mbox{   ,   } \nonumber \\
&& \Omega_{\rm i}(\zeta,\theta) = \omega_{\rm i}^{(0)}(\theta) + \omega_{\rm i}^{(1)}(\theta) e^{-\zeta} \mbox{   ,   }
\end{eqnarray}
where the index ``i'' denotes that this set of solutions is only valid 
in the inner boundary layer. The integrating functions are related 
to one another by 
\begin{equation}
s_{\rm i}^{(1)}(\theta) = \left(\frac{4\pi \rho\nu}{\eta}\right)^{1/2} \ri^2 \sin^2\theta  \omega_{\rm i}^{(1)}(\theta) \mbox{   .   }
\label{eq:rel2}
\end{equation}

In the case of the boundary conditions 
selected here, the solution in the bulk of the fluid is well-approximated by 
neglecting any effect of dissipation. This result was formally 
shown by Dormy, Cardin \& Jault (1998). 
There, $S$ and $\Omega$ are constant along magnetic field 
lines as discussed earlier, so we can write the bulk solution as 
\begin{equation}
S_{\rm b}(r,\theta) = s_{\rm b}\left(A(r,\theta)\right) \mbox{   and   } \Omega_{\rm b}(r,\theta)= \omega_{\rm b}\left(A(r,\theta)\right)\mbox{   .   }
\end{equation}

As $\xi$ and $\zeta$ respectively tend to $+\infty$, the boundary 
solutions must approach the bulk solution smoothly. Hence,
\begin{eqnarray}
s_{\rm i}^{(0)}(\theta) = s_{\rm b}\left(A(\ri,\theta)\right) \mbox{   and   } 
s_{\rm o}^{(0)}(\theta) = s_{\rm b}\left(A(\ro,\theta)\right) \mbox{   ,   } \nonumber \\
\omega_{\rm i}^{(0)}(\theta) = \omega_{\rm b}\left(A(\ri,\theta)\right)\mbox{   and   } 
\omega_{\rm o}^{(0)}(\theta) = \omega_{\rm b}\left(A(\ro,\theta)\right)\mbox{   .   }
\end{eqnarray}
Since the co-latitudes $\thetai$ and $\thetao$ satisfy by definition
$A(\ri,\thetai) = A(\ro,\thetao)$, the above relationships can 
be summarized as
\begin{equation}
s_{\rm i}^{(0)}(\thetai) = s_{\rm o}^{(0)}(\thetao) \mbox{   and   } \omega_{\rm i}^{(0)}(\thetai) = \omega_{\rm o}^{(0)}(\thetao)\mbox{   .   }
\label{eq:eqset1}
\end{equation}

Finally, we apply the boundary conditions to determine the unknown integration functions uniquely. Requiring (\ref{eq:ominbc}) and (\ref{eq:omczbc}) implies that
\begin{eqnarray}
\omega_{\rm i}^{(0)}(\theta) + \omega_{\rm i}^{(1)}(\theta) = \Omega_{\rm in} \mbox{   ,   }\nonumber \\
\omega_{\rm o}^{(0)}(\theta) + \omega_{\rm o}^{(1)}(\theta) = \Omega_{\rm cz}(\theta)\mbox{   .   }
\label{eq:eqset2}
\end{eqnarray}
Requiring $S=0$ on both boundaries implies 
\begin{eqnarray}
s_{\rm i}^{(0)}(\theta) + s_{\rm i}^{(1)}(\theta) = 0 \mbox{   ,   }\nonumber \\
s_{\rm o}^{(0)}(\theta) + s_{\rm o}^{(1)}(\theta) = 0\mbox{   .   }
\label{eq:eqset3}
\end{eqnarray}

The set of eight equations contained in (\ref{eq:rel1}), (\ref{eq:rel2}), 
(\ref{eq:eqset1}), (\ref{eq:eqset1}) and  (\ref{eq:eqset3}) can be solved 
uniquely for the eight integration functions. Of particular interest for 
the following calculation is the expression for $\omega_{\rm i}^{(1)}$:
\begin{equation}
\omega_{\rm i}^{(1)}(\thetai) = - \frac{\ro^3}{\ro^3 - \ri^3} \left( \Omega_{\rm cz}(\thetao)  -\Omega_{\rm in} \right)\mbox{   ,   }
\label{eq:omegai1}
\end{equation}
where $\thetai$ and $\thetao$ are related by equation (\ref{eq:thetaio}).

We now seek to express $\Omega_{\rm in}$ as a weighted integral over
$\Omega_{\rm cz}(\theta)$. Applying (\ref{eq:notorque}) on the inner boundary, 
and using (\ref{eq:zetadef}) and (\ref{eq:omegai1}) yields
\begin{equation}
\int_0^{\pi/2} \sin^3 \theta  \frac{\partial \Omega_{\rm i}}{\partial r } \dd \theta = - \int_0^{\theta_{\cal C}} \frac{\sin^3 \thetai}{\deltai f(\thetai)}  \omega_{\rm i}^{(1)}(\thetai) \dd \thetai = 0\mbox{   ,   }
\end{equation}
since, in the equatorial region on the inner core ($\thetai > \theta_{\cal C}$), 
$\partial \Omega /\partial r = 0$. Changing variables from $\thetai$ to 
$\thetao$, and using the actual expression for $f(\theta)$ 
transforms this equation to 
\begin{equation}
\int_0^{\pi/2} \cos \thetao \sin^3 \thetao \left[ \Omega_{\rm cz}(\thetao) - \Omega_{\rm in} \right] \dd \thetao = 0\mbox{   ,   }
\end{equation}
which implies that 
\begin{equation}
\Omega_{\rm in} = \Omega_{\rm eq} \left(1 - \frac{a_2}{3} - \frac{a_4}{6}\right) \mbox{   .   }
\end{equation}
Using the helioseismically determined values for $a_2$ and $a_4$ yields
\begin{equation}
\Omega_{\rm in} = 0.93 \Omega_{\rm eq}\mbox{   .   }
\end{equation}
In the asymptotic limit where this expression is valid 
(e.g. $\deltai/\ri \ll 1$) note that $\Omega_{\rm in}$ is independent 
of the aspect ratio $\ri/\ro$ of the setup, and can be shown to hold for 
non-homogeneous fluids as well. In fact, it only depends 
on the imposed differential rotation.

\subsubsection{Numerical solutions}

We obtained numerical solutions to the set of linearized equations 
(\ref{eq:indphi}) and (\ref{eq:NSphi}) with 
boundary conditions (\ref{eq:ominbc}), (\ref{eq:omczbc}) and (\ref{eq:s0bc})
using a numerical algorithm adapted from the one developed by Garaud (2001). 
This algorithm involves the expansion of the governing equations upon a 
truncated set of Chebishev functions in $\mu = \cos\theta$, and seeks the 
solution of the 
remaining set of ODEs by Newton-Raphson relaxation. The linear nature
of the equations (in the variables $S$ and $\Omega$ considered) 
guarantees the immediate convergence of the solutions.

The analytical results presented above are asymptotically 
independent of the individual 
values selected for $B_0$, $\eta$ and $\rho \nu$, and depend instead only 
on their combination through the Hartman number $H$, which is the ratio
of the geometric mean of the two diffusion times (viscous and magnetic) 
to the Alfv\'en time. From here on, we define the Hartman number of 
the simulation as 
\begin{equation}
H = \frac{B_0 \ri}{\sqrt{ 4\pi \rho \nu \eta}} \mbox{   .   }
\end{equation}

\begin{figure}[h]
\epsscale{1}
\plotone{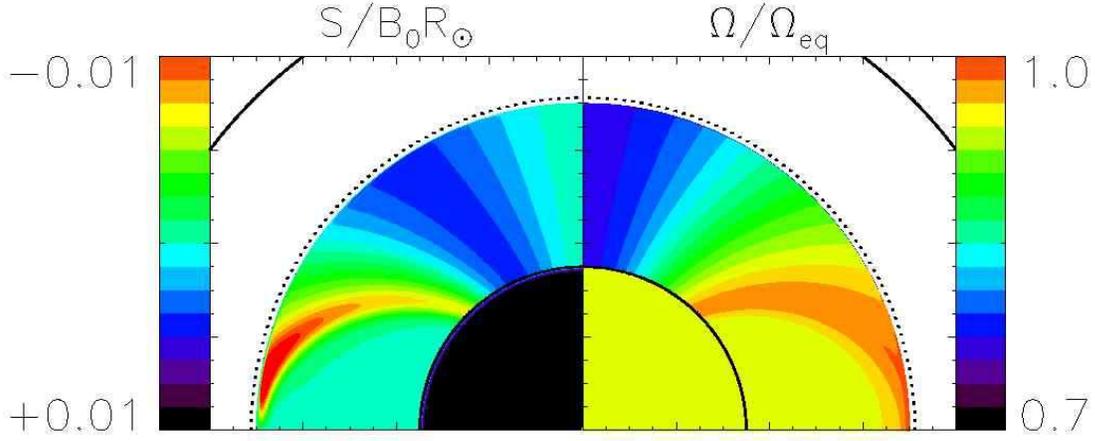}
\caption{Unconfined field solution for $H = 4700$ and for $S=0$ on both boundaries. In our attempt to relate these simulations to the solar radiative zone, the outer solid circle marks for reference the outer edge of the Sun ($r = R_\odot$), and the dotted line marks the radiative--convective interface. Here, $\ro = 0.7 R_\odot$ and $\ri = 0.35 R_\odot$.}
\label{fig:unconfs0}
\end{figure}

Figure \ref{fig:unconfs0} shows the numerical solution for a fairly large value of the Hartman number, $H = 4700$. As expected, the equatorial region is indeed rotating with the same angular velocity as the core, and $S$ is zero there. In the polar regions, both $\Omega$ and $S$ are constant along poloidal magnetic field lines, except near the inner and outer boundaries, and in the vicinity of the field line $({\cal C})$. The rotation rate of the inner core is found to be $\Omega_{\rm in} = 0.937 \Omega_{\rm eq}$ in this simulation. This numerical solution confirms our analytical results. Moreover, as $H \rightarrow +\infty$, we find that $\Omega_{\rm in} \rightarrow 0.98 \Omega_{\rm eq}$ (see Figure \ref{fig:omvsf}).

\subsection{Conducting boundary conditions}
\label{sec:conducting}

We now turn to the case of ``conducting boundary conditions'' for the
toroidal field, which are the same conditions as those used in the
numerical simulations of the radiative zone dynamics performed by 
GG08. The material outside of the fluid
shell is assumed to be infinite in extent and 
to have the same diffusivity as the fluid inside the 
shell. The toroidal field satisfies the equation 
\begin{equation}
\nabla^2 B_\phi - \frac{B_\phi}{r^2 \sin^2\theta} = 0 \mbox{   ,   }
\end{equation}
for $r > \ro $ and $r < \ri$, with $B_\phi \rightarrow 0$ as 
$r\rightarrow 0$ and $r \rightarrow + \infty$. The solution to this 
equation is smoothly matched onto the solution within the spherical shell 
at $r=\ri $ and $r = \ro$ (see GG08, for detail).

As first shown by Hollerbach (2000) the characteristics of the 
solution are now different from the case of insulating boundary 
conditions, an effect which is studied in detail by Soward \& Dormy (2009).
Remarkably, the bulk of the fluid is not in Ferraro isorotation. 
To understand the difference on a qualitative basis, note that
the magnetic field is merely diffusing out of the boundaries and that 
the continuous generation of toroidal field by the fluid motions within
the shell is only compensated by this dissipation. In other words, nothing
but dissipation limits the growth of the amplitude of the field. 
As a result, the amplitude of $S$ is proportional to 1/$\eta$, so 
that the approximation $\bBp \cdot \nabla \Omega = 0$ in the bulk of the fluid 
is no longer valid (the diffusion term is $O(1)$ compared with the 
advection term). Note that $\bBp \cdot \nabla S = 0$ still holds. 

\begin{figure}[h]
\epsscale{1}
\plotone{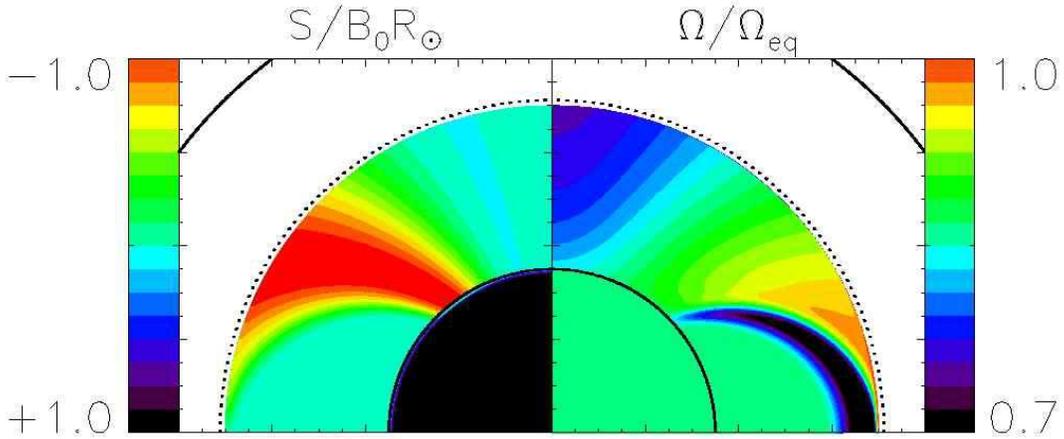}
\caption{Same as Figure \ref{fig:unconfs0} but for conducting boundary conditions (using $(\nabla^2 \bB)_{\phi}=0$ for $r > \ro$ and $r < \ri$). Note how the 
angular velocity profile is not in a state of Ferraro rotation; also note the
difference in the amplitude of $S$. }
\label{fig:unconfgrad}
\end{figure}

Since the bulk equations are notably more difficult to solve analytically
in this case (see Soward \& Dormy 2009 for detail), 
we only provide a sample numerical solution in Figure 
\ref{fig:unconfgrad}. The parameters for this simulation are exactly the same
as for the case shown in the previous section -- only the magnetic field 
boundary  conditions differ. The two characteristic features mentioned
above are clearly illustrated in Figure \ref{fig:unconfgrad}: (1) the amplitude
of the $S$ field is orders of magnitude larger than before, although 
$S$ is still constant on poloidal magnetic field lines; (2) the angular 
velocity profile deviates from Ferraro isorotation. A strong sub-rotating
layer appears just interior to the field line (${\cal C}$), and the angular 
velocity of the inner core $\Omega_{\rm in} = 0.863 \Omega_{\rm eq}$ 
is much slower than in the previous calculation (\S3.1). Finally, note that the 
core velocity found in this case does depend on the aspect ratio of the system. The variation of $\Omega_{\rm in}$ with $H$ is shown in Figure \ref{fig:omvsf}.

\section{Solution for a confined field}
\label{sec:closed}

In order to represent a confined magnetic field, we select the flux 
function $A(r,\theta)$ as in  the work of R\"udiger \& Kitchatinov (1997) 
to be
\begin{equation}
A(r, \theta) = B_0 \frac{r^2}{2} \left( 1- \frac{r}{\ro} \right)^q \sin^2 \theta \mbox{   ,   }
\label{eq:Aconf}
\end{equation}
where $q$ is the confinement parameter, which we assume is greater or 
equal to one. Note that when $q=1$, $B_r = 0$ at $\ro$ while $B_\theta$ 
is non zero. On the other hand, for $q > 1$, $B_\theta$ also vanishes 
at $r = \ro$. Also note that the constant 
$B_0$ now defines the amplitude of the magnetic field at $r=0$ (i.e. not 
on the polar axis at $r=\ri$).

\begin{figure}[h]
\epsscale{1}
\plotone{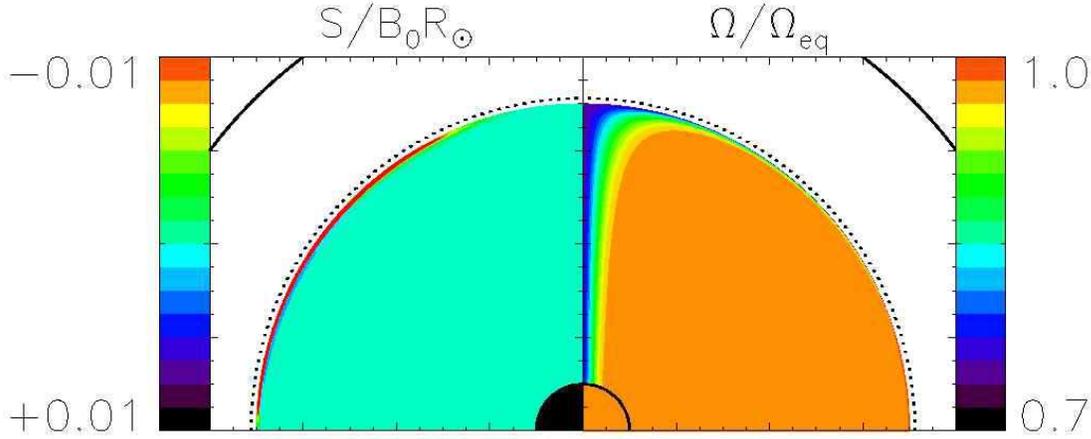}
\caption{Numerical solution in the confined field case for $q=1$ and $H=1340$, using insulating boundary conditions. Here, $\ri = 0.1$ and $\ro = 0.7$. The angular velocity is constant in 
the bulk of the ``radiative zone'', with $\Omega_{\rm in} = 
0.972\Omega_{\rm eq}$.}
\label{fig:confd}
\end{figure}

We begin by considering the case of insulating boundary conditions, 
where $B_\phi = 0$ at $\ri$ and $\ro$. Figure \ref{fig:confd} presents a 
numerical solution for large Hartman number, for $q=1$. It clearly shows that
the bulk of the fluid is rotating uniformly with the same angular velocity 
as the inner core, and that the azimuthal magnetic field is zero in the same 
region. This bulk solution is expected on the 
same symmetry grounds as the ones 
invoked in the equatorial region of the 
unconfined field case. It matches smoothly onto the applied boundary 
conditions at $r = \ro$ 
through a boundary layer which is very thin everywhere except near the polar
axis. 

Following the method used in the case of the unconfined field configuration, 
we begin by rescaling the radial variable near $\ro$ with 
\begin{equation}
\xi = \frac{\ro - r}{\deltao f(\theta)} \mbox{   ,   }
\end{equation}
where the typical layer width $\deltao$ and the form function $f(\theta)$ 
remain to be determined. As before, we substitute this new variable 
for $r$ in equations (\ref{eq:indphi}) and (\ref{eq:NSphi}) 
and keep the lowest order terms in $\deltao$ only, so that 
\begin{eqnarray}
&& \frac{\eta \ro^{q-2}}{B_0 \deltao^{q+1}} 
 \xi^{1-q} \frac{\partial^2 S}{\partial \xi^2} 
= f^{q+1}(\theta)\sin^2\theta  \left[  \frac{q \sin\theta}{2} \frac{\partial \Omega}{\partial \theta} + \left( \frac{q \sin\theta}{2} \frac{f'(\theta)}{f(\theta)}  +  \cos\theta \right)  \xi \frac{\partial \Omega}{\partial \xi} \right]   \mbox{   ,   }\nonumber \\
&& \frac{4\pi\rho\nu \ro^{q+2}}{B_0  \deltao^{q+1}} 
\xi^{1-q} \frac{\partial^2 \Omega}{\partial \xi^2} 
= \frac{ f^{q+1}(\theta) }{\sin^2\theta} \left[  \frac{q \sin\theta}{2} \frac{\partial S}{\partial \theta} + \left( \frac{ q \sin\theta}{2} \frac{f'(\theta)}{f(\theta)}  +  \cos\theta \right)  \xi \frac{\partial S}{\partial \xi} \right]   \mbox{   .   }
\label{eq:bleqsconf}
\end{eqnarray}
If we select
\begin{equation}
\deltao^{2q+2} = \frac{16 \pi \rho \nu \eta \ro^{2q} }{q^2 B_0^2}  \mbox{   and   } f(\theta) = \frac{1}{\sin^{2/q} \theta}\mbox{   ,   }
\label{eq:blpredconf}
\end{equation}
then the variables fully separate, and the boundary layer equations become
\begin{eqnarray}
&&  \xi^{1-q}\frac{\partial^2}{\partial \xi^2} \left( \xi^{1-q}\frac{\partial^2 S}{\partial \xi^2} \right) = (1-\mu^2)^{1-1/q} \frac{\partial }{\partial \mu} \left[ (1-\mu^2)^{-1-1/q} \frac{\partial S}{\partial \mu} \right] \mbox{   ,   } \nonumber \\
&&  \xi^{1-q}\frac{\partial^2}{\partial \xi^2} \left( \xi^{1-q}\frac{\partial^2 \Omega}{\partial \xi^2} \right) = (1-\mu^2)^{-1-1/q} \frac{\partial}{\partial \mu}\left[ (1-\mu^2)^{1-1/q} \frac{\partial \Omega}{\partial \mu} \right]\mbox{   ,   }
\label{eq:nastybleqs}
\end{eqnarray}
with the introduction of the variable $\mu = \cos\theta$. 

\begin{figure}[h]
\epsscale{0.5}
\plotone{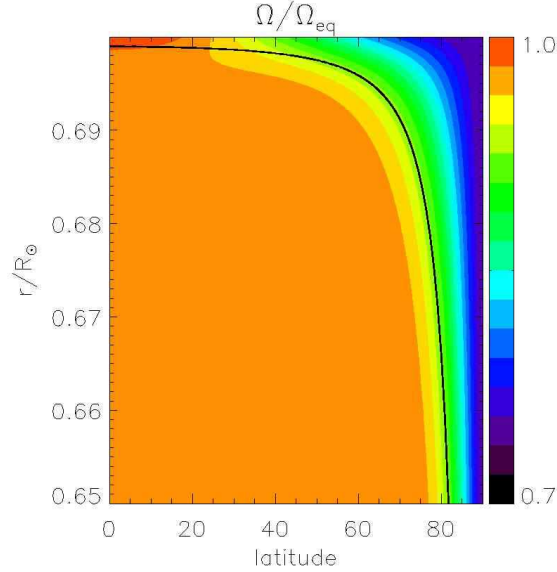}
\caption{Close up of the angular velocity profile near the outer 
boundary layer for $q=1$ and $H=1340$ (see Figure \ref{fig:confd}).
Superimposed on this numerical solution we show the predicted boundary 
layer thickness $\deltao f(\theta)$ divided by three 
for the same parameter values -- the two are in excellent agreement. }
\label{fig:confcomp}
\end{figure}

We validate our boundary layer approximation by comparing, in 
Figure \ref{fig:confcomp}, the predicted latitudinal structure of the 
boundary layer for $q=1$ with numerical solutions. The agreement is 
excellent, and 
similar comparisons for other values of $q$ are also in excellent
agreement with (\ref{eq:blpredconf}).

\begin{figure}[h]
\epsscale{0.5}
\plotone{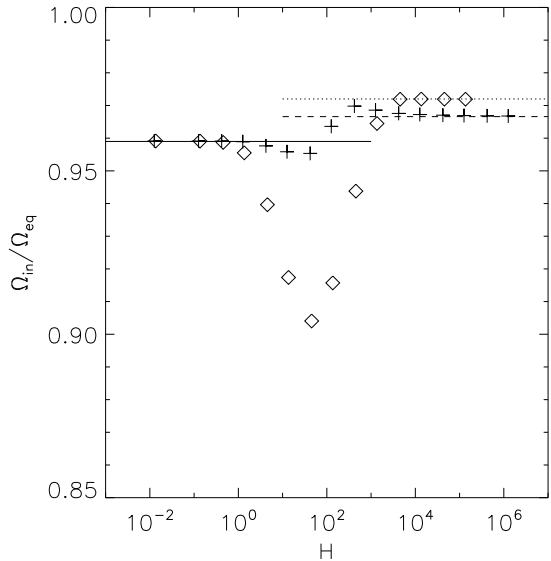}
\caption{Calculated core velocity as a function of Hartman number. This figure compares numerical simulations and analytical predictions for $\Omega_{\rm in}/\Omega_{\rm eq}$ for two values of the confinement parameter $q$, for insulating boundary conditions. The diamonds show numerical results for $q=1$, and the plus symbols for $q=2$. The dotted line marks the analytical asymptotic limit for $q=1$ and the dashed line for $q=2$ (see equation ({eq:ominconf2})). In the low Hartman number limit on the other hand, all curves converge to the value $\Omega_{\rm visc}/\Omega_{\rm eq}$ (solid line).}
\label{fig:omvsfconf}
\end{figure}

Even when $q=1$, 
solving (\ref{eq:nastybleqs}) analytically is not entirely trivial\footnote{An approximate solution can be found for $q=1$.}. 
Luckily, estimating the value of the interior angular velocity does not
require knowledge of the full boundary layer solution. We can obtain a first-order accurate approximation of the derivative $\partial \Omega / \partial r $ across the boundary layer with 
\begin{equation}
\left. \frac{\partial \Omega}{\partial r}\right|_{(\ro,\theta) }\simeq \frac{\Omega_{\rm cz}(\theta) - \Omega_{\rm in} }{\deltao f(\theta)} \mbox{   ,   } \label{eq:blextapprox}
\end{equation}
for large Hartman number. Applying (\ref{eq:notorque}) at $\ro$ and using the approximation (\ref{eq:blextapprox}), we find that the angular velocity of the interior is determined by 
\begin{equation}
\int_0^1 (1-\mu^2)^{1+1/q} \left[ \Omega_{\rm cz}(\mu) - \Omega_{\rm in} \right] \dd \mu = 0\mbox{   ,   }
\end{equation}
which implies
\begin{equation}
\Omega_{\rm in} = \Omega_{\rm eq} \left( 1 - \frac{a_2}{5 + 2/q} - \frac{3a_4}{(7+2/q)(5+2/q)} \right)  \mbox{   .   }
\label{eq:ominconf1}
\end{equation}
Note that as $q\rightarrow \infty$ (when the field is more and more confined),
the solution recovers the purely viscous case as expected. 
For the helioseismically determined values of $a_2$ and $a_4$ we get
\begin{eqnarray}
\Omega_{\rm in} \simeq 0.972 \Omega_{\rm eq}\mbox{    for   } q = 1  \mbox{   ,   } \nonumber \\ 
\Omega_{\rm in} \simeq 0.966 \Omega_{\rm eq}\mbox{    for   } q = 2\mbox{   ,   }
\label{eq:ominconf2}
\end{eqnarray}
and so forth. 
To verify our analysis, we compare the 
asymptotic values of $\Omega_{\rm in}$ calculated in equation (\ref{eq:ominconf1}) with the numerical solutions for different values of $q$ in Fig. \ref{fig:omvsfconf}. The agreement, for high values of the Hartman number, is again excellent. Finally, it can be shown that by contrast with 
the unconfined field case, changing boundary conditions for the toroidal field
does not yield a different asymptotic answer for $\Omega_{\rm in}$. This can be attributed to the fact that the bulk solution, and the geometry of the boundary layer, are the same regardless of the boundary conditions. 

\section{Discussion}
\label{sec:disc}

In the two previous sections, we studied the axially symmetric, linearized
dynamics of homogeneous magnetized spherical Couette flows for various 
geometries of the imposed poloidal field and for various types of magnetic 
boundary conditions. We now discuss how these simplified models may help 
us make sense of the dynamics of the solar radiative zone and the tachocline. 

\subsection{Comparison with the numerical simulations of GG08}
\label{sec:disc1}

Recently, GG08 presented a set of numerical simulations of the solar 
radiative zone and the tachocline based on the Gough \& McIntyre model (GM98). 
By contrast with the toy model studied above, GG08 attempt to 
model the dynamics of the radiative interior as accurately as possible 
and solve the following system of equations
\begin{eqnarray}
&& \overline{\rho} \tilde{\bu} \cdot \grad \tilde{\bu} + 2 \overline{\rho} {\bf \Omega}_0 \times \tilde{\bu}   =
- \grad \tilde{p} - \tilde{\rho} \grad \overline{\Phi}
+ \bj \times \bB + f_\nu \div \Pi \mbox{   ,   }\nonumber  \\
&&  \div(\overline{\rho} \tilde{\bu}) = 0 \mbox{   ,   } \nonumber  \\
&& \overline{\rho} \overline{T} \tilde{\bu}\cdot \grad \overline{s}
= \div( f_k \overline{k} \grad \tilde{T})\mbox{   ,   } \nonumber  \\
&& \frac{\tilde{p}}{\overline{p}} = \frac{\tilde{\rho}}{\overline{\rho}} +  \frac{\tilde{T}}{\overline{T}}\mbox{   ,   }\nonumber \\
&& \curl(\tilde{\bu} \times \bB) = \curl( f_\eta \overline{\eta} \curl \bB) \mbox{   ,
  }
\nonumber  \\
&& \div \bB = 0\mbox{   ,   }
\label{eq:global2}
\end{eqnarray}
for the velocity field in
the rotating frame $\tilde{\bu} = (\tilde{u}_r,\tilde{u}_\theta,r \sin\theta \tilde \Omega)$, for the magnetic field $\bB = (B_r,B_\theta, B_\phi)$ and for the 
density ($\tilde{\rho}$), temperature ($\tilde{T}$) and pressure ($\tilde{p}$) perturbations. In these equations,
${\bf \Omega}_0$ is equal to $\Omega_{\rm visc}$ (see equation (\ref{eq:ovisc})), 
the background stratification for $\bar \rho$ (the density), 
$\bar T$ (the temperature), $\bar s$ (the entropy), $\bar \Phi$ (the 
gravitational potential) and $\bar p$ (the pressure) 
are given by Model S of Christensen-Dalsgaard {\it et al.} (1996), and the diffusion coefficients $\bar \nu$ (for the viscosity), $\bar \eta$ (for the 
magnetic diffusivity) and $\bar k$ (for the thermal conductivity) are calculated according to Gough (2007) (see also GG08). Large factors $f_\nu$, $f_\eta$ and $f_k$ multiply these respective diffusivities to help numerical convergence. The boundary conditions used on the inner core are:
\begin{itemize}
\item impermeable and no-slip, with $\tilde\Omega = \Omega_{\rm in} - \Omega_0$ and where $\Omega_{\rm in}$ is deduced from (\ref{eq:notorque}),
\item electrically conducting for the electric/magnetic field 
(e.g. the magnetic field satisfies
$\grad^2 \bB = 0$ in the inner core, matches on to a point dipole at $r=0$,
and matches continuously to the solution in the spherical shell at $r=\ri$),  
\item thermally conducting for the temperature field (e.g. $\tilde{T}$ 
satisfies $\grad^2 \tilde{T} = 0$ in the inner core, and smoothly matches on
to the solution in the shell at $r = \ri$).
\end{itemize}
The boundary conditions used on the outer boundary are:
\begin{itemize}
\item the velocity field matches smoothly onto an imposed velocity field
$$\bu^{\rm cz} = \left(u_r^{\rm cz}, u^{\rm cz}_{\theta}, \ro \sin\theta (\Omega_{\rm cz}-\Omega_0)\right),$$
\item electrically conducting for the electric/magnetic field 
(e.g. the magnetic field satisfies
$\grad^2 \bB = 0$ for $r > \ro$, vanishes as $r \rightarrow \infty $ and
smoothly matches onto the shell solution at the boundary, 
\item thermally conducting for the temperature field (e.g. $\tilde{T}$ 
satisfies $\grad^2 \tilde{T} = 0$ for $r>\ro$, and smoothly matches on
to the solution in the shell at $\ro$.
\end{itemize}

The numerical solutions of these equations and boundary conditions, as computed 
by GG08, exhibit many of the dynamical properties of the tachocline first 
discussed by GM98. In order to prevent the propagation
of the convection zone shear into the interior (as illustrated in Figures 
\ref{fig:unconfs0} and \ref{fig:unconfgrad}), confining the primordial field
to the radiative zone appears to be necessary. GM98 argued that 
large-scale meridional flows downwelling from the convection zone would
naturally interact nonlinearly with the underlying field and confine 
it beneath the bulk of the tachocline. The 
convection zone flows are by the same mechanism prevented from penetrating 
more deeply into the radiative zone, thereby satisfying a variety of 
observational constraints on chemical mixing near the base of the 
convection zone (e.g. Elliott \& Gough 1999). The tachocline would 
thus be a well-ventilated region, spatially separated from the 
magnetically-dominated interior by a very thin advection-diffusion layer.
All of these features are qualitatively well-accounted for in the simulations
of GG08. 

However, the angular velocity of the bulk of the radiative interior in 
the simulations of GG08 is much lower than the observed value: 
$\Omega_{\rm in} \simeq 0.87 \Omega_{\rm eq}$ in the limit of large Hartman 
number. Surprisingly, the same value is found for a fairly wide range 
of assumed convection zone flow amplitudes and profiles $u_r^{\rm cz}(\theta)$. Given that simulations robustly insist on selecting 
this particular interior angular velocity, can we understand it in terms of 
the simple dynamics studied in the toy model? Surprisingly, it appears that 
we can. 

\begin{figure}[ht]
\epsscale{0.5}
\plotone{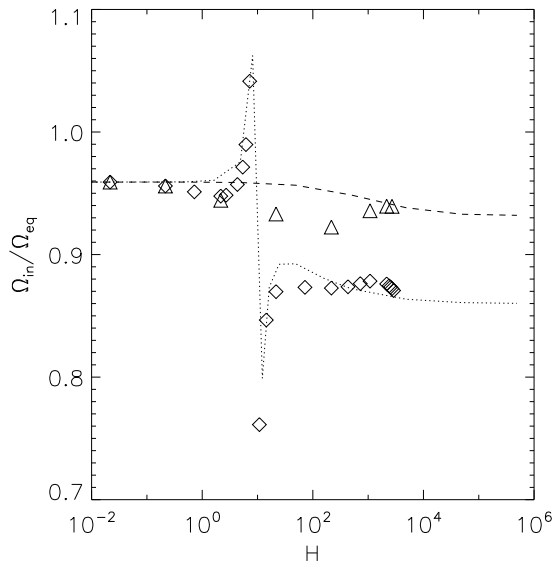}
\caption{Calculated angular velocity of the core as a function of Hartman number. This figure compares the linear solutions in the toy-model (lines) and fully nonlinear numerical solutions (symbols) for two different types of boundary conditions. The conducting boundary conditions, shown in the dotted line and diamond symbols, were also used by GG08. In the limit of large Hartman number, the interior rotation rate $\Omega_{\rm in}$ tends to about $0.87 \Omega_{\rm eq}$. In the second case, shown with the dashed line and triangles, we changed the lower boundary to be insulating. The asymptotic limit in both toy model and in the full numerical simulations then appears to be closer to the observations, with $\Omega_{\rm in} \rightarrow 0.93 \Omega_{\rm eq}$.}
\label{fig:omvsf}
\end{figure}

To see this quantitatively, Figure \ref{fig:omvsf} compares
the core angular velocity predictions for the stratified, nonlinear 
calculations of GG08 with our toy-model calculations for an open dipole
in a similar experimental setup (i.e. the same aspect ratio and with conducting 
boundary conditions). The agreement between the two sets of simulations 
as a function of Hartman number is quite remarkable, in spite of the 
over-simplified nature of the toy model. 

We did in 
fact {\it expect} the agreement to be very good for high values of the 
diffusivities (low values of $H$). Indeed, since the background is 
strongly stratified, only very slow meridional flows can penetrate 
into the radiative interior (see Garaud \& Brummell 2008). 
The magnetic Reynolds number of these flows 
is therefore also low, and they fail to have any influence on 
the poloidal field. The magnetic field then naturally relaxes to its
fundamental eigenmode, which is the open dipole. These 
combined factors together imply that the linearized problem 
studied in \S\ref{sec:open} should be (and is indeed found to be) 
a good approximation to the overall angular momentum balance of the system.
It correctly predicts the asymptotic limit $H \rightarrow 0$, as
well as the somewhat surprising bifurcation around $H = 1$.
It is also interesting to note that radial variations in $\rho$, $\nu$ 
and $\eta$ do not appear to affect the predictions for the interior angular 
velocity much. This can be formally shown in the case of insulating
boundary conditions but not in the case of conducting boundary conditions. 
Nevertheless, it appears that the agreement approximately holds for 
conducting boundary conditions as well. 

The good agreement between the predicted core angular velocities in the 
toy model and in GG08's simulations, for low values of the diffusivities 
(high values
of $H$), is much more surprising. The lowest-diffusivities 
simulations presented by GG08, which correspond to the right-most symbols
in Figure \ref{fig:omvsf} (see also Figure \ref{fig:fullsims}), 
have a magnetic field geometry which 
deviates significantly from the purely dipolar open-field 
configuration studied in \S\ref{sec:open}. 
Moreover, the same simulations clearly show 
the existence of a region where angular-momentum 
transport is operated by the meridional flows rather than by 
the magnetic field, as predicted by GM98.
Both phenomena are a natural consequence of the increasingly nonlinear nature
of the interaction between the primordial field and the assumed 
convection zone flows as the diffusivities are lowered (equivalently, 
as $H$ is increased); both should by and large invalidate the applicability
of the toy model. Nevertheless, even then we find that the open-dipole 
toy model 
adequately predicts the core angular velocity of the fully nonlinear numerical 
simulations, for the parameter values considered. 

Whether this statement would continue to hold for even lower values of the 
diffusivities (even higher values of $H$) is {\it a priori} 
unlikely\footnote{although cannot be ruled out.}. 
As the importance of angular-momentum
transport by the meridional flows increases in relation to viscous transport, 
we expect significant deviations away from the toy model predictions to occur. 
Indeed, while GG08 failed to achieve low enough diffusion parameters in their
simulations to test this hypothesis, they also presented another
set of simulations for artificially high convection zone flow amplitudes 
(see their Figure 11), in which the calculated core velocities do 
deviate significantly away from the toy-model predictions.
Moreover, one could also expect that as the 
magnetic field becomes even more confined to the interior, the open-dipole
configuration will lose relevance in favor of the closed-dipole configuration.
Since we have shown that the closed-dipole predictions 
are closer to $\Omega_{\rm in} \sim 0.97 \Omega_{\rm eq}$, we may expect the 
predicted core velocity in the full simulations to increase towards this 
value as the diffusivities are decreased\footnote{unless angular-momentum 
transport by the flows acts just in the opposite way, see footnote 2.}.

\subsection{Sensitivity to boundary conditions}
\label{sec:disc2}

The core velocity found in the linearized model is
sensitively dependent on the assumed magnetic field boundary conditions. 
This was shown in \S\ref{sec:open}, where changing the 
boundary conditions from insulating to conducting had a profound 
impact on the nature of the solution. In fact, it can also be 
shown that the same happens by changing from conducting conditions to having 
even just one boundary condition where $B_\phi=0$ (see below). 
But far more importantly, 
this sensitive dependence on boundary conditions is {\it also} found in the 
numerical solutions of the full set of equations (\ref{eq:global2}). 

\begin{figure}[ht]
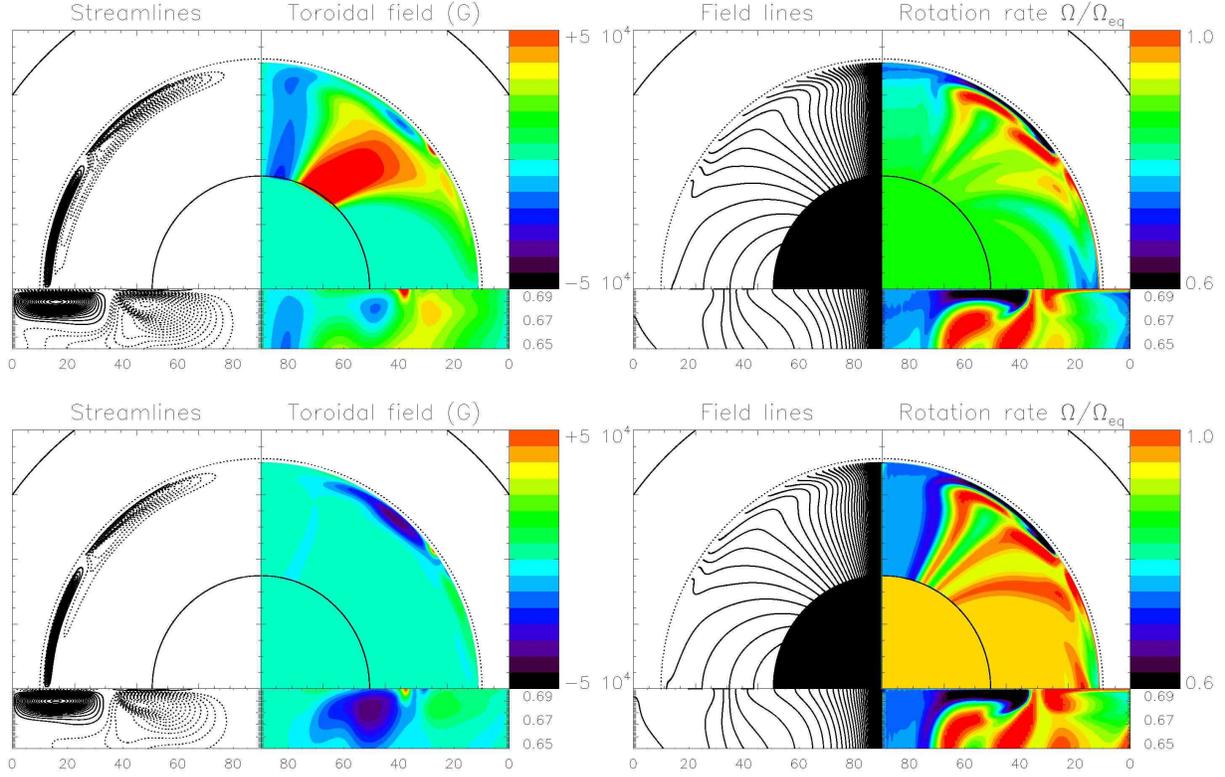

\epsscale{1}
\plotone{f8a.epsf}
\plotone{f8b.epsf}
\caption{Comparison between the numerical solutions of (\ref{eq:global2}) with two different sets of boundary conditions on the toroidal field. {\it Top:} Conducting boundary conditions everywhere. {\it Bottom:} The top boundary is conducting while the bottom boundary is insulating. All other simulation parameters are exactly the same for the two runs: $B_0 = 7$T, $f_\nu = 8 \times 10^8$, $f_\kappa = 8 \times 10^7$ and $f_\eta = 8 \times 10^9$. The strip beneath each quadrant zooms into the region near the outer boundary, for $r \in [0.65,0.7] R_\odot$. The numbers represent latitude. In the streamlines panel, solid lines denote clockwise flows and dotted lines anti-clockwise flows.} 
\label{fig:fullsims}
\end{figure}

We ran a separate set of simulations for both the linearized toy 
model and for the full nonlinear model, where the outer boundary 
is again conducting but the inner core is now insulating 
(in the sense that $B_\phi = 0$ at $r = \ri$). The predictions for the 
core angular velocity as a function of the Hartman number are shown in 
Figure \ref{fig:omvsf}. We found that in the large-Hartman number limit, 
the ``asymptotic'' value of $\Omega_{\rm in}$ in the full nonlinear model
is much closer to the observed value, $0.93 \Omega_{\rm eq}$, as also predicted
by the toy model. Figure \ref{fig:fullsims} compares the numerical 
results of the full 
simulations with the different sets of boundary conditions with 
one-another. While the streamlines and the poloidal
field lines (as well as the temperature and density profiles, not shown) 
appear to be relatively unaffected by the
new boundary condition at $\ri$, 
the angular velocity and toroidal field profiles
are notably different. Following the trends of the toy model,  
the amplitude of the toroidal field is 
significantly lower and the rotation profile is much closer 
to Ferraro isorotation when at least one of the boundary conditions is not 
conducting.

The implications of these findings are quite important. Short of simulating
the entire solar interior including the turbulent convection zone and its 
effect on magnetic fields, one needs to make assumptions on the nature
of the radiative--convective interface. It appears that models in which
the fluid shell is contained in a conducting solid of 
infinite extent are somewhat
pathological in nature, as they allow an 
unphysically high magnetic field amplitude to build up thus breaking 
away from Ferraro isorotation (Soward \& Dormy 2009). 
Meanwhile, insulating boundary conditions seem
to be an {\it a priori} equally poor physical representation of both 
the inner core and of the 
radiative--convective interface. In reality, one may either expect 
other physical phenomena to limit the toroidal field amplitude 
within the radiative zone (e.g. magnetic instabilities), or at the very least 
note that the 
Sun is not infinite in extent, so that $B_{\phi}= 0$ in the vacuum outside
of $R_\odot$. So, as strange as it may seem, having at least one insulating
boundary may actually be more physically realistic than the boundary conditions
originally used by GG08.

\subsection{Implications for models and observations of the solar interior}
\label{sec:disc3}

Aside from the demonstrably odd case of the conducting boundary conditions 
described in the previous section, all simple analytical models of the 
radiative zone presented so far predict an angular velocity $\Omega_{\rm in}$ close to the observations: in all cases (see Table 1), $0.908 \Omega_{\rm eq} \le \Omega_{\rm in} \le 0.972 \Omega_{\rm eq}$. Crucially, all of the estimates presented in Table 1 are independent of $\rho$, $\nu$, $\eta$, $B_0$ and of the aspect ratio $\ro/\ri$, in the asymptotic limit of large Hartman number. In addition, Garaud (2002) also 
studied the hydrodynamic case of spherical Couette flow between one differentially and one uniformly rotating sphere, and showed that when the gap width is 
of the order of the observed thickness of the tachocline, the predicted
angular velocity of the interior is also, perhaps coincidentally, close to
the observed value of $ \Omega_{\rm rz} = 0.93 \Omega_{\rm eq}$.

\begin{table}
\begin{center}
\caption{Summary of analytical model predictions}
\begin{tabular}{ccc}
\tableline\tableline
Model type & $\Omega_{\rm rz}/\Omega_{\rm eq}$  & $\Omega_{\rm rz}/\Omega_{\rm eq}$\tablenotemark{a} \\
\tableline
Viscous model\tablenotemark{b} & $ 1 - \frac{a_2}{5} - \frac{3a_4}{35}$ & 0.959 \\
Anisotropic viscosity\tablenotemark{c} &$ 1 - \frac{3a_2}{7} - \frac{5a_4}{21}$ & 0.908 \\
Open dipole field\tablenotemark{d} &  $1 - \frac{a_2}{3} - \frac{a_4}{6}$ & 0.930  \\
Confined dipole field\tablenotemark{e} & $1 - \frac{a_2}{ 5 + 2/q} - \frac{3a_4}{ (5+2/q)(7+2/q) } $   & 0.959 - 0.972  \\
\tableline
\end{tabular}
\tablenotetext{a}{Using $a_2=0.17$ and $a_4=0.08$, Schou {\it et al.} (1998)}
\tablenotetext{b}{Gilman, Morrow \& DeLuca (1989)}
\tablenotetext{c}{Spiegel \& Zahn (1992)}
\tablenotetext{d}{see \S\ref{sec:insulating}}
\tablenotetext{e}{see \S\ref{sec:closed}}
\end{center}
\end{table}

Table 1 has a somewhat ironic
property: the models which are a priori the most unphysical, or the poorest
representation of the solar interior are the ones which actually seem to fare
the best in terms of predicting $\Omega_{\rm in}$ close to the observed value.
Indeed, recall that 
the open-dipole case has a non-uniformly rotating radiative zone, while the 
hydrodynamic spherical Couette flow (Garaud, 2002) assumes the fluid to be
confined between two impermeable spherical shells. 

There are several lessons to be learned from this work. 
As mentioned in \S1, 
all predictions for $\Omega_{\rm in}$ necessarily involve a weighted integral
over $\Omega_{\rm cz}(\theta)$. Moreover, the spherical geometry of the
problem implies that the weight function is typically biased towards 
the equatorial regions -- in other words, $\Omega_{\rm in}$ is  
more sensitive to $a_2$ than to $a_4$. As a result, we see that the spread
in predictions for $\Omega_{\rm in}$ is relatively small, and one should
neither be surprised to see many different models predicting similar values, 
nor that some should lie coincidentally close to the observed one. 

Nevertheless, it is equally interesting to see that $\Omega_{\rm in}$ 
in the closed-dipole model, which is perhaps the ``closest'' 
(in relative terms) 
to what one may expect from the tachocline dynamics, {\it is} significantly 
different from the observations. This implies one of two things: either 
meridional flows (or perhaps anisotropic turbulent stresses) are a 
non-negligible contribution to angular-momentum transport in the tachocline
or (if they are negligible) the true 
angular velocity profile near the base of the convection zone deviates
significantly away from the one used here (see equation (\ref{eq:ocz})). 
Helioseismology may be able to help distinguish between these two alternatives.

\section{Conclusion}
\label{sec:ccl}

We have studied, analytically and numerically, the predicted angular
velocity profile of the solar radiative zone under various 
model assumptions. Our overall conclusions have implications for future 
modeling, and implications for future observations.  

In terms of modeling, we have illustrated how crucial the selection of 
magnetic boundary conditions can be to the calculated solution, an effect
which has only recently been fully appreciated (see the detailed study by 
Soward \& Dormy, 2009). Assuming, as previous models have done 
(Garaud, 2002; Brun \& Zahn 2006; GG08), that the radiative zone is contained 
within a homogeneous conducting medium of infinite extent allows unphysically 
large toroidal field amplitude to build up. This case is a somewhat
pathological limit, since if the toroidal field is somehow forced to be zero 
at a finite radius (e.g. the solar photosphere), or if other mechanisms act 
to limit its amplitude, then the problem does not arise. Nevertheless, how
to {\it best} represent the presence of the solar convection zone remains to 
be determined.

In terms of observations, our various calculations have quantified the 
sensitivity of the angular velocity of the interior to the model assumptions:
aside from a few exceptional cases which can be ruled out (see above)
the predicted angular velocity lies roughly in the interval
$[0.91 \Omega_{\rm eq},0.97 \Omega_{\rm eq}]$. Angular-momentum 
balance between viscous stresses and magnetic stresses for a closed-dipole 
suggests that $\Omega_{\rm in} \simeq 0.97 \Omega_{\rm eq}$. If helioseismic
observations can rule out this value entirely, then we can conclude from this 
study that the tachocline is the seat of additional mixing, either in the 
form of large-scale meridional flows (Gough \& McIntyre, 1998), or 
in the form of small-scale turbulence. Although chemical evidence for 
additional mixing in the tachocline has already been
put forward (Gough \& McIntyre, 1998; Elliott \& Gough, 1999, R\"udiger \& Pipin, 2001), our work provides the first dynamical evidence to this effect. 

\section*{Acknowledgements}

This work originated from C. Guervilly's summer project at the Woods
Hole GFD Summer School in 2008. We thank the NSF and the ONR for supporting
this excellent program. P. Garaud was supported by NSF-AST-0607495. The
numerical simulations were performed on the Pleiades cluster at UCSC, 
purchased using an NSF-MRI grant. We thank L. Acevedo-Arreguin, 
N. Brummell, G. Glatzmaier, 
D. Gough, T. Wood and the Woods Hole GFD staff for many fruitful discussions.

\end{document}